\documentstyle[12pt]{article}

\def\half{{1\over2}}

\begin{document}
\thispagestyle{empty}
{\baselineskip=12pt
\hfill IFP-9901-UNC 

%\hfill hep-th/98

\hfill January 1999 

\vspace{1.0cm}}
\centerline{\large \bf Symmetric Subgroups of Gauged Supergravities} 
\centerline{{\large \bf and AdS String Theory Vertex Operators}
\footnote{Work supported in part by the U.S. Department of Energy
under Grant No. DE-FG05-85ER40219/Task A.}}
\bigskip
\bigskip
\centerline{\large L. Dolan\footnote{Email: dolan@physics.unc.edu} 
and M. Langham\footnote{Email: mlangham@physics.unc.edu}}
\medskip
\centerline{\it Department of Physics and Astronomy}
\centerline{\it University of North Carolina, Chapel Hill, NC 27599-3255}
\bigskip
\bigskip
\bigskip

\vskip 1.0 truein
\parindent=1 cm

\begin{abstract}
We show how the  gauge symmetry representations of the
massless particle content of gauged supergravities that arise in the
AdS/CFT correspondences can be derived from {\it symmetric subgroups} 
to be carried by string theory vertex operators in these compactified models,
although an explicit vertex operator construction of IIB string 
and M theories on AdSxS remains elusive.
Our symmetry mechanism parallels the construction  
of representations of the Monster group and affine algebras 
in terms of twisted conformal field theories, and may serve as a guide
to the perturbative description of the IIB string on AdSxS.

\end{abstract}
\vfil\eject
\setcounter{page}{1}
\section{Introduction}

Recently gauged supergravities have become natural candidates
for low-energy limits of superstring and M theories on anti-de Sitter space
times a sphere (AdSxS)\cite{mal,kleb,wi1,wi2}. This was somewhat unexpected 
since conformal invariance of string theory
implies a vanishing tree level cosmological constant for the higher-dimensional
theory. Although sigma model descriptions have been found\cite{tsy}, 
the precise formulation of these theories via vertex operators, {\it i.e.}
primary fields, has remained elusive\cite{gks}. In this paper, we 
assign representations of a symmetric subalgebra of the gauge symmetry
of the supergravity to left and right-moving vertex operators, and show
that the full gauge group representations are then regained by 
inverse branching rules. This mechanism is similar to the 
construction of affine algebras and representations of the
Monster group as twisted conformal field theories\cite{flm,dgm, ch}.
The vertex operators 
associated with the gauge bosons in the supergravity multiplet thus 
include both bosonic and fermionic emission operators.
Our identification of the subgroups may serve as a guide in the explicit
construction of these vertex operators, and the subsequent description
of IIB on AdSxS as a perturbative string theory. 

In sect.'s 2-4,
we compute symmetric subgroups for the
three examples that have maximal supersymmetry.
Each corresponds to a collection of $p$-branes 
whose near horizon geometry
looks like $AdS_{p+2}\times S^{{\cal D}-p-2}$, where ${\cal D} =$ 10 or 11
for branes in string or M-theory, and $p=2,3,5$.
The formulation of vertex operators and
string theory tree amplitudes on $AdS_5$ will allow access
to the dual conformal $SU(N)$ gauge field theory $CFT_4$ at large $N$, but
{\it small} fixed `t Hooft coupling $x = g^2_{YM} N$ in the dual 
correspondence, as 
$(g^2_{YM} N)^\half (4\pi)^\half = R^2_{sph}/{\alpha'}$.
Presently only the large $N$, and {\it large} fixed `t Hooft coupling $x$
limit is accessible in the CFT, since only the supergravity limit
($\alpha'\rightarrow 0$) of the AdS theory is known\cite{mf}.
In sect. 5 we discuss a case with non-maximal supersymmetry.

\section{IIB Superstring on $AdS_5\times S^5$}

For $p=3$,  a set of 
$D3$-branes in type IIB superstring theory has near horizon
geometry $AdS_5\times S^5$.  
The low-energy limit is $D=5$, $N=8$ gauged supergravity~\cite{grw,krv,gm} 
and the physical massless states belong to the supermultiplet with
$(1; 4 +\bar 4; 15 + 6 +6; 4 + \bar 4 + 20 + \bar{20}; 20_R + 10 + \bar{10}
+ 1 + 1)$ 
representations of the gauge group SU(4).
The last set of representations
correspond to the 42 scalars which parameterize the coset space $E_{6,6}/Sp_4$. 

The ungauged $D=5$, $N=8$ supergravity (corresponding to the low enery limit
of the IIB superstring on ${\rm Minkowski}_5\times T^5$) has the same number of 
physical states as the gauged version, but 
with light-cone little group Spin(3) and particle content
$({\underline 5}, 8\, {\underline 4}, 
27\, {\underline 3}, 48\, {\underline 2},
42\, {\underline 1})$ derived below. (Spin(3) representations are 
underlined). Unlike the gauged version, this theory can be 
easily constructed from vertex operators given in terms of free worldsheet
boson and fermion Fock space operators. Working in the Neveu-Schwarz
Ramond formalism in light-cone gauge, we find the massless sector as follows.
The left and right-moving modes are each
taken to be described by eight bosonic and eight
fermionic worldsheet fields
~\cite{gsw,bdg}.
Both left and right-moving 
fermionic (Neveu-Schwarz(NS)) fields are in the ${\underline 3} + 
5\,{\underline 1}$ representation of Spin(3). 
Spin fields, corresponding to massless Ramond (R) states, transform
as $4\,{\underline 2}$ of Spin(3) on each side. The (left x right)
massless states of $D=5$, $N=8$ are given by:

\noindent NS-NS \hskip5pt $({\underline 3} + 5\, {\underline 1})
\times ({\underline 3} + 5\, {\underline 1}) = 
{\underline 5} + 11\, {\underline 3} + 26\,{\underline 1}$;\,

\noindent R-R \hskip14pt $4\,{\underline 2}\times 4\,{\underline 2} =
16\, ({\underline 3} + {\underline 1} )$

\noindent NS-R \hskip5pt $({\underline 3} + 5\,{\underline 1})\times
4\,{\underline 2} = 4\, {\underline 4} + 24\, {\underline 2}$	

\noindent R-NS \hskip5pt $({\underline 3} + 5\,{\underline 1})\times
4\,{\underline 2} = 4\, {\underline 4} + 24\, {\underline 2}$. 

For the gauged version, the construction of vertex operators for the 
underlying type IIB string theory on $AdS_5\times S^5$ remains obscure,
but as in superstring theory on flat space, it seems reasonable that
the physical string states in this maximally symmetric theory on 
anti-de Sitter space are also in one-to-one correspondence with 
vertex operators. We show that if vertex operators
with the same number of
spin degrees of freedom as in the flat space case, are in the
following representations of $SU(2)^2\times U(1)$, a symmetric subgroup
of the gauge group $SU(4)$, then the resulting closed string massless states 
can be identified with Table 1. 

The procedure is thus to identify the left and right-moving modes
as certain representations of a symmetric subgroup of the supergravity
gauge group. There will be massless gauge bosons from both the 
NS-NS and the R-R sectors. Gauge bosons corresponding to the adjoint
represention of the symmetric subgroup are in the NS-NS sector.
The R-R bosons carry a representation of the subgroup, and together
with the NS-NS bosons they enhance the symmetry to 
the adjoint representation of the
full gauge group. This mechanism is familiar from conformal field theory
constructions of twisted affine algebras and the Monster 
group~\cite{flm,dgm,ch}. The bosons from the twisted (Ramond) sector
form part of the non-abelian multiplet. 

A symmetric subgroup has the property that
its structure constants can be written as 
$f_{ABC} = \{f_{abc}, f_{aIJ}, {\rm otherwise}\, 0\}$, for
$1\le A\le$ dimension of the gauge group, $1\le a\le$
dimension of the subgroup, and $A=\{ a, I\}$. See for eg. ref [16]. 
The fact that the subgroup we choose is a symmetric subgroup, where 
$a$ labels states in the NS sectors and $I$ states in the Ramond
sectors, leads to a correlation between allowed trilinear string state
couplings (ie NS-NS-NS, or NS-R-R, but not NS-NS-R nor R-R-R) and
the gauge couplings ($f_{abc}$ or $f_{aIJ}$ non-zero, but $f_{abI}=f_{IJK}=0$).
This feature holds for our three examples.  

Note that although there is the same number of physical degrees
of freedom in the `massless' sector of the ungauged and gauged models,
the light-cone little group analysis is different, and the elimination
of non-physical degrees of freedom proceeds differently in the
two cases~\cite{af, krv, sez, ffz}. Nonetheless, for convenience in the tables,
we list the gauged supergravity states by their ``Spin'', 
{\it i.e.} what would be
the light-cone little group representation in the ungauged version.
The $AdS$ vertex operators carrying these quantum numbers would thus
be in ``light-cone'' gauge and describe physical degrees of freedom.
They would be in one to one correspondence with Hilbert space states given 
by unitary short representations of the supergroup
$SU(2,2/4)$, and with the non-physical degrees freedom subtracted out.
The supergroup for this model, $SU(2,2/4)$, has 30 even generators
in the subgroup $SU(2,2)\times SU(4)$, and 32 odd generators.
The even subgroup is the isometries of $AdS_5$ and $S^5$ respectively,
where the algebras $su(2,2)\sim so(2,4)$ and $su(4)\sim so(6)$.

For the type IIB superstring on $AdS_5\times S^5$, we assign the
following $SU(2)^2\times U(1)$ representations to the left and 
right-moving NS modes and spin fields (the numbers in the
(,) brackets are the $SU(2)^2$ representations, the subscripts are
$U(1)$ helicity, and the $D=5$ multiplicity appears underlined preceeding 
each bracket):
\vskip5pt

\noindent NS \hskip6pt (left)\hskip7pt {\underline 3} \,$(1,1)_0,$\quad
{\underline 1} \,$((3,1)_0 + (1,1)_2 + (1,1)_{-2})$ 

\noindent NS (right)\hskip7pt {\underline 3} \,$(1,1)_0,$\quad
{\underline 1}\, $((1,3)_0 + (1,1)_2 + (1,1)_{-2})$

\noindent Ramond \hskip6pt (left)\hskip7pt {\underline 2}\, $((2,1)_1 +
(2,1)_{-1})$

\noindent Ramond (right)\hskip7pt {\underline 2}\, $((1,2)_1 + (1,2)_{-1})$.
\vskip5pt

Taking the left $\times$ right tensor products of these 
$SU(2)^2$ representations for the NS-NS, R-R, NS-R, R-NS sectors,
we then reproduce the values in the last column of Table 1. 
Since these representations are precisely those in the branching rules
of the $SU(4)$ representations in the second column, we conjecture that
vertex operators carrying such quantum numbers will reproduce the
string theory with $D=5, N=8$, SU(4) gauged supergravity. 
Although at this time a vertex operator construction for
the type IIB superstring on $AdS_5\times S^5$ is not known, 
it seems likely that the theory has
a perturbative conformal field theory description and therefore
that the string states including the low-energy supergravity fields
will have a vertex operator description. Our analysis presents the
representation theory content for such operators.  

\begin{table}[t]
\caption{The massless D=5, N=8 gauged supergravity
multiplet.\label{tab:exp1}}
\vspace{0.2cm}
\begin{center}
\scriptsize
\begin{tabular}{@{}|c@{}|@{}c@{}|@{}c@{}|@{}l@{}|}
\hline
{``Spin(3)''} &
\raisebox{0pt}[13pt][7pt]{SU(4) rep} &{Dynkin label}
& \raisebox{0pt}[13pt][7pt]{Branching rules to $SU(2)^2
\times U(1)$}\\
\hline
&&&\\
$5$  &
1 & (000) & $(1,1)_0$ \\[5pt]
$4$   &  
$4 + \bar 4$& (100) + (001) & 
$(2,1)_1 + (1,2)_{-1}$ \\
&&&+$(2,1)_{-1} + (1,2)_{1}$ \\[5pt]
$3$  &
15  + 6 + 6
& (101)  + (010) + (010)&  
$(1,1)_0 + (3,1)_0 + (1,3)_0$ \\
&&&+$(2,2)_2 + (2,2)_{-2}$\\
&&&+$2(1,1)_2 + 2(1,1)_{-2} + 2(2,2)_0$\\[5pt]
$2$     &
$4 + \bar 4 + 20 + \bar {20}$& (100)  + (001) + (011) + (110)& 
$(2,1)_1 + (2,1)_{-1} + (1,2)_{-1} + (1,2)_1$\\
&&&$+(3,2)_{-1} + (2,3)_1$\\
&&&$+ (2,1)_{-3} + (2,1)_1 + (1,2)_{-1} + (1,2)_3$\\
&&&$+(3,2)_1 + (2,3)_{-1}$\\
&&&$+ (2,1)_3 +(2,1)_{-1} + (1,2)_1 + (1,2)_{-3}$\\[5pt]
$1$   &  
$1+1+10+\bar{10} +20_R$ &  (000) + (000) + (200) + (002) + (020)
& $ 2 (1,1)_0 $\\
&&&+ $(3,1)_2 + (1,3)_{-2} + (2,2)_0$\\
&&&+ $(3,1)_{-2} + (1,3)_2 + (2,2)_0$\\
&&&+ $(3,3)_0 + (2,2)_2 + (2,2)_{-2}$\\
&&&+ $(1,1)_0 + (1,1)_4 + (1,1)_{-4}$\\
\hline
\end{tabular}
\end{center}
\end{table}

\section{M-theory on $AdS_4\times S^7$}

For $p=2$, a set of M2-branes in M-theory has near horizon
geometry $AdS_4\times S^7$. 
The low-energy limit is $D=4$, $N=8$ gauged
supergravity~\cite{dwn,dnp,gw} and the
massless states have physical degrees of freedom given by the familiar
supermultiplet with 
$(1; 8_s; 28; 56_s; 35_v + 35_c)$
representations of the gauge group SO(8).
Here the last set of representations corresponds to the 70 scalars
which parameterize the coset space $E_{7,7}/SU(8)$.
The supergroup is $OSp(8/4)$ which has 38 even generators in the subgroup
$SO(2,3)\times SO(8)$, and 32 odd generators.

We assign the following $SU(2)^4$ representations to the left and
right-moving NS modes and spin fields where the ( , , , ) bracket labels 
$SU(2)^4$, and the first bracket labels the ``Spin(2)'' as in the
first column of Table 2:
\vskip5pt

\noindent NS \hskip6pt (left)\hskip7pt $(\pm1)$\, (1,1,1,1), \quad
(0)\, ( (3,1,1,1) + (1,3,1,1) )

\noindent NS (right)\hskip7pt $(\pm1)$\, (1,1,1,1), \quad
(0)\, ( (1,1,3,1) + (1,1,1,3) )

\noindent Ramond \hskip6pt (left)\hskip7pt $(\pm 1/2)$ (2,2,1,1)

\noindent Ramond (right)\hskip7pt $(\pm 1/2)$ (1,1,2,2).

Computing the left $\times$ right tensor products of these
$SU(2)^4$ representations for the NS-NS, R-R, NS-R, R-NS sectors,
we reproduce the values in the last column of Table 2,
and conjecture that vertex operators in light-cone gauge
carrying such quantum numbers will be useful in contructing 
string theory with $D=4, N=8$, SO(8) gauged supergravity as its
low-energy limit ~\cite{dolan}.

\begin{table}[t]
\caption{The massless D=4, N=8 gauged supergravity 
multiplet.\label{tab:exp2}}
\vspace{0.2cm}
\begin{center}
\footnotesize
\begin{tabular}{|c|c|c|l|}
\hline
{``Spin(2)''} &
\raisebox{0pt}[13pt][7pt]{SO(8) rep} &{Dynkin label}
&\raisebox{0pt}[13pt][7pt]{Branching rules to $SU(2)^4$}\\
\hline
&&&\\
$\pm 2$   &
1 & (0000)& (1,1,1,1)\\[5pt]
$\pm {3/2}$   & 
$8_s $& (0001) & (2,2,1,1) + (1,1,2,2)\\[5pt]
$\pm 1$   &  
28 & (0100) &(3,1,1,1) + (1,3,1,1)\\
&&&+(1,1,3,1) + (1,1,1,3)\\
&&&+ (2,2,2,2)\\[5pt]
$\pm 1/2$   &  
$56_s$& (1010) & (2,2,1,1) + (1,1,2,2)\\
&&&+(2,2,3,1) + (2,2,1,3)\\
&&&+(3,1,2,2)+(1,3,2,2)\\[5pt]
$0$   &
$35_v + 35_c$ &  (2000) + (0020)
&(3,1,3,1) + (1,3,1,3) \\
&&&+(1,3,3,1) + (3,1,1,3)\\
&&&+ 2(2,2,2,2) + 2(1,1,1,1)\\[5pt]
\hline
\end{tabular}
\end{center}
\end{table}

\section{M-theory on $AdS_7\times S^4$}

For $p=5$, a set of 5-branes in M-theory has near horizon
geometry $AdS_7\times S^4$. 
The low-energy limit is $D=7$, $N=4$
gauged supergravity~\cite{ppn,gnw,ptn,vn}
and the massless supergravity multiplet has
$(1; 4; 5; 10; 16; 14)$
representations of the gauge group SO(5). The 14 scalars 
parameterize the coset space $SL(5,R)/SO(5)$.
The supergroup is $OSp(2,6/4)$ which has 38 even generators in the 
subgroup $SO(2,6)\times SO(5)$, and 32 odd generators.
This is the simplest of our three examples.
We choose the $su(2)^2$ symmetric subalgebra of $so(5)$ as listed in
Table 4 and
assign the following $SU(2)^2$ representations to the
left and right-moving NS modes and spin fields:
\vskip5pt

\noindent NS \hskip4pt (left)\hskip7pt {\underline 5}\, (1,1),\quad
{\underline 1}\, (3,1)

\noindent NS (right)\hskip7pt {\underline 5} \,(1,1),\quad
{\underline 1}\, (1,3)

\noindent Ramond \hskip4pt (left)\hskip7pt {\underline 4}\, (2,1)

\noindent Ramond (right)\hskip7pt {\underline 4}\, (1,2).
\vskip5pt

The tensor products of the $SU(2)^2$ representations which then result 
in the NS-NS, R-R, NS-R, R-NS sectors are computed in the last column of
Table 3. These representations are those which occur in the
branching rules for $SO(5)\supset SU(2)^2$ and give the representations in
the second column that are found in the $D=7$, $N=8$ SO(5) gauged supergravity  
multiplet. The properties of the structure constants in a symmetric
subalgebra ensure that gauge couplings of the states are consistent
with string tree amplitudes for NS and Ramond states.

\begin{table}[t]
\caption{The massless D=7, N=4 gauged supergravity multiplet.\label{tab:exp3}}
\vspace{0.2cm}
\begin{center}
\footnotesize
\begin{tabular}{|c|c|c|l|}
\hline
{``Spin(5)''} & 
\raisebox{0pt}[13pt][7pt]{SO(5) rep} &{Dynkin label}
&\raisebox{0pt}[13pt][7pt]{Branching rules to $SU(2)
\times $SU(2)} \\
\hline
&&&\\
$14$   & 
1 & (00)& (1,1) \\
$16$ &
$4$ & (10) & (2,1) + (1,2) \\
$10$   & 
$5 $& (01) & (1,1) + (2,2)\\
$5$  & 
10& (20) & (1,3) + (3,1) + (2,2)\\
$4$ &
$16$ &  (11) & (2,1) + (1,2) + (3,2) + (2,3)\\
$1$ & 
14 & (02) & (1,1) + (2,2) + (3,3)\\[5pt]
\hline
\end{tabular}
\end{center}
\end{table}

\begin{table}[t]
\caption{Symmetric subgroups.\label{tab:exp4}}
\vspace{0.2cm}
\begin{center}
\footnotesize
\begin{tabular}{|c|c|c|l|}
\hline
{$p$} &\raisebox{0pt}[13pt][7pt]{Low energy theory} &
\raisebox{0pt}[13pt][7pt]{Supergravity gauge group} &{Symmetric subgroup}\\
\hline 
&&&\\
$2$ & $D=4, N=8$ & $SO(8)$ & $SU(2)^4$\\
$3$ & $D=5, N=8$ & $SU(4)$ & $SU(2)^2\times U(1)$\\
$5$ & $D=7, N=4$ & $SO(5)$ & $SU(2)^2$\\[5pt]
\hline
\end{tabular}
\end{center}
\end{table}

\section{IIB Superstring on $AdS_3\times S^3$}

Finally, we consider the non-maximally supersymmetric case
of the IIB string on $AdS_3\times S^3\times M^4$, where 
$M^4$ is either $T^4$ or $K3$~\cite{gks, dkss, jdb, lar, kll, jms, bort, ggt}. 
The supergroup for this model is $SU(2/1,1)\times SU(2/1,1)$, which 
has 12 even generators in the subgroup $SU(1,1)\times SU(1,1)\times
SO(4)$, and 16 odd generators. $SU(1,1)^2$ is the 
isometry group of $AdS_3$ and $SO(4)$ that of $S^3$. 
The low-energy limit is 
$D=3$, $N=8$, $SO(4)$ gauged supergravity.
In three dimensions the graviton and gravitinos do not have physical 
degrees of freedom, and the physical spectrum consists of 
$n$ copies of eight bosons and eight fermions 
in representations of the gauge group $SO(4)$ given by 
$n\,((2,2) + 4(1,1);  2(1,2)+2(2,1))$, where
$n=5$ for $M^4=T^4$ and $n=21$ for $M^4=K3$.

Because both
$AdS_3$ and $S^3$  are group manifolds, a vertex operator construction
is straightforward in this theory. $AdS_3$ corresponds to the 
$SU(1,1)$ manifold, and $S^3$ is the $SU(2)$ manifold. 
The gauge symmetry algebra 
$so(4)\sim su(2)^2$ thus appears explicitly as left and right worldsheet 
$su(2)$ currents, and 
in this $AdS_3$ case 
there is no symmetric subalgebra construction. That is to say,
the vertex operators carry quantum numbers of the full gauge symmetry 
group, even before taking the left times right tensor products. So no new
gauge symmetry generators appear in the Ramond sector in this
non-maximally supersymmetric example, in contrast to our suggestion
for the maximally symmetric cases discussed in sect.'s 2-4. 

\section{Conclusions}

In the three maximally symmetric examples of the AdS/CFT correspondence,
we have analyzed the representations of the gauged supergravities
in terms of symmetric subalgebras. Although each of the
gauge algebras in our models, 
ie. $su(4)\sim so(6)$, so(8) and $so(5)\sim sp_2$, 
has several symmetric subalgebras,
we identify a particular one (for each case) that appears to have
a natural role in terms of the quantum numbers of left and right-moving
vertex operators. In the case of $AdS_5$ ($SU(4)$), our symmetric subgroup
$SU(2)\times SU(2)\times U(1)$ coincides
with the subgroup with respect to which $SU(4)$ has Jordan structure
used ~\cite{gm, gs} in the oscillator method to construct unitary irreducible 
representations of the AdS theory. A direct connection between this latter
construction and a vertex operator construction, however, is not apparent. 

The enhancement of gauge symmetry from the subalgebra that
occurs as a current algebra in each sector is familiar from the
conformal field theory construction of affine algebras and sporadic
groups by twisted
{\it i.e.} orbifold-like conformal field theories. 
For example, for an $E_8$ affine algebra, the untwisted and
twisted sectors each carry a representation of the subgroup
$SO(16)$, but together the two sectors can be combined by an inverse
branching ratio to form $E_8$ representations, with some of the
states in the adjoint representation of $E_8$ coming from the twisted sector. 
In this case, the currents will consist of both superfields and spin fields.

\end{document}